\documentclass[showpacs,amsmath,amssymb,aps,showkeys,floatfix,prd,a4paper]{revtex4}
\usepackage{subfigure}
\usepackage{dcolumn}
\usepackage{bm}
\usepackage{epsfig}
\usepackage{amsfonts}
\usepackage{amssymb,amscd}
\usepackage{soul}
\usepackage{color}

\newcommand{\be}{\begin{equation}}
\newcommand{\ee}{\end{equation}}
\def\beq{\begin{equation}}
\def\eeq{\end{equation}}
\def\beqa{\begin{eqnarray}}
\def\eeqa{\end{eqnarray}}
\newcommand{\ba}{\begin{eqnarray}}

\begin{document}



\title{Production of meson molecules in ultra-peripheral heavy ion collisons}

\author{F.C. Sobrinho$^{1}$, L.M. Abreu$^{1,2}$, C.A. Bertulani$^{3,4}$, 
F.S. Navarra$^{1}$\\
$^1$Instituto de F\'{\i}sica, Universidade de S\~{a}o Paulo,
Rua do Mat\~ao 1371 - CEP 05508-090,
Cidade Universit\'aria, S\~{a}o Paulo, SP, Brazil\\
$^2$Instituto de F\'{\i}sica, Universidade Federal da Bahia,
Campus Ondina, Salvador, Bahia 40170-115, Brazil\\
$^3$Department of Physics and Astronomy, Texas A\&M University-Commerce,
Commerce, Texas 75429, USA\\
$^4$Institut f\"ur Kernphysik,  
Technische Universit\"at Darmstadt, 64289 Darmstadt, Germany\\
}


\begin{abstract}

In this work we present the first calculation of exotic charmonium production 
in ultra-peripheral collisions, in which the exotic state is explicitly 
treated as a meson molecule. Our formalism is general but we focus on the 
lightest possible exotic charmonium state: a $D^+ D^-$ molecular bound state.
It was proposed some time ago and it has been object of experimental 
searches. Here we study the production of the 
open charm pair in the process $\gamma \gamma \to D^+ D^-$. Then we use a 
prescription to project the free pair $ |D^+ D^- \rangle$ onto a bound 
state at the amplitude level and compute the cross section of the process
$\gamma \gamma \to B$ (where $B$ is the bound state). Finally, we convolute 
this last cross section with the equivalent photon distributions coming from 
the projectile and target in an ultra-peripheral collision and find the 
$A A \to A A B$ cross section, which, for $Pb-Pb$ collisions at 
$\sqrt{s_{NN}} = 5.02$ TeV, is of the order of $3 \, \mu \mbox{b}$.

\end{abstract}




\maketitle

\vspace{0.5cm}


\section{Introduction}

One of the most important research topics in modern hadron physics is the 
study of the  exotic heavy quarkonium states \cite{exo,exo2}. These new mesonic 
states are not conventional $c \bar{c}$ configurations and their minimum 
quark content is  $c \bar{c} q \bar{q}$. This leads us to the main question 
in the field: are these multiquark states compact tetraquarks or are they 
meson molecules? So far there is no conclusive answer. One can try to 
address this question with the help of experiment and study the observables:
masses, decay widths and production rates. How can multiquark states be 
produced? In $B$ decays and in $e^+ e^-$, proton-proton, proton-nucleus and    
nucleus-nucleus collisions. We will focus on the latter, which can be divided 
into central (and semi-central) and ultra-peripheral (UPCs) \cite{BKN05}.    
In UPCs the nuclei do not overlap and there are only few particles produced. 
In these collisions the elementary processes which contribute to particle 
production are photon-photon, photon-Pomeron and Pomeron-Pomeron fusion. 
The advantage of UPCs is the low particle-production multiplicity, thus with 
a reduced background if proper detection techniques are used. Such features 
have been explored at the large hadron collider (LHC) at CERN and at the 
relativistic heavy ion collider (RHIC) at Brookhaven.

In this work we will study exotic charmonium production in photon-photon 
processes in nucleus-nucleus collisions. Coming back to the question formulated
above, the strategy to get the answer is to compute the cross section for 
production (in UPCs) of a given exotic charmonium state assuming that it is 
i) a tetraquark and also assuming that it is ii) a meson molecule. We believe  
that the resulting cross sections are very different from each other and 
hence, just looking at the production rate, one could experimentally 
discriminate between the two configurations. Here we will address only the 
production of molecules. The study of tetraquark production is in progress.

The production of hadron molecules has been discussed in the context of $B$  
decays \cite{pi}, in $e^+ e^-$ collisions,  in proton-proton      
\cite{pp1,pp2}, in proton-nucleus and in central nucleus-nucleus 
collisions \cite{aa}. In this work we present the first 
study of meson molecule production in UPCs. The method employed here is 
applicable to all molecular states. We start with the lightest charm meson 
molecule: the $D^+ D^-$ state (also called $D\bar{D}$). 
It was predicted in the study of meson-meson 
interactions in the charm sector in \cite{15}, where it was found to be 
bound by 
about 20 MeV. The state was confirmed in subsequent theoretical studies 
\cite{16,17}. More recently it was also found in lattice calculations 
\cite{18}. 
In \cite{19}, it was shown that the peak in the $D \bar{D}$ invariant mass, 
observed by the BELLE collaboration \cite{be}, could be well explained by
the existence of a hidden charm scalar resonance below the threshold 
\cite{15}. An updated experimental work was performed in 
\cite{20} and, again, support for the $D \bar{D}$ state in the reaction 
$e^+ e^- \to D \bar{D}$ (and also in $\gamma \gamma \to D \bar{D}$) was found. 
Recent analyses of these data were published in \cite{21,22}. 
A more refined theoretical work of these reactions was performed in      
\cite{mb,23,ao}, claiming again evidence for this bound state. 

In the next Section we present the formalism employed to describe $D^+ D^-$ 
pair production; in Section III we present the prescription to create the 
bound state; in Section IV we discuss the equivalent photon spectrum; in 
Section V, performing a low energy approximation, we derive an analytical 
formula for the cross section of bound state production. In the final 
section we present numerical results and discussion.

\section{Production of free $D^+D^-$ pairs}

There are two ways to produce a $D^+ D^-$ from two photons. In the first, one 
of the photons splits directly into the pair $\gamma \to D^+ D^-$, where one 
of the mesons is already on the mass shell, and the second photon brings 
the other $D$ to the mass shell. This process can be described by a well known 
hadronic effective Lagrangian, from which we obtain the pair production 
amplitude. This amplitude is subsequently projected onto the amplitude for 
bound state formation. If the properties of the bound state are known, the   
only unknown in this formalism is the form factor, which must be attached to 
the vertices to account for the finite size of the hadrons. 

In the second way to produce the pair, one photon splits into a $c \bar{c}$ 
pair which, after interacting with the second photon, hadronizes into the 
$D^+ D^-$ pair. Then, using  a coalescence prescription, we obtain a model 
for the production of the bound
state. The hadronization process involves uncertainties related to its 
non-perturbative nature. Here we can not automatically use fragmentation 
functions, which require a hard scale. Moreover, the coalescence prescription 
contains some inherent arbitrariness. 

\begin{figure}
    \centering
    \includegraphics[width=.60\linewidth]{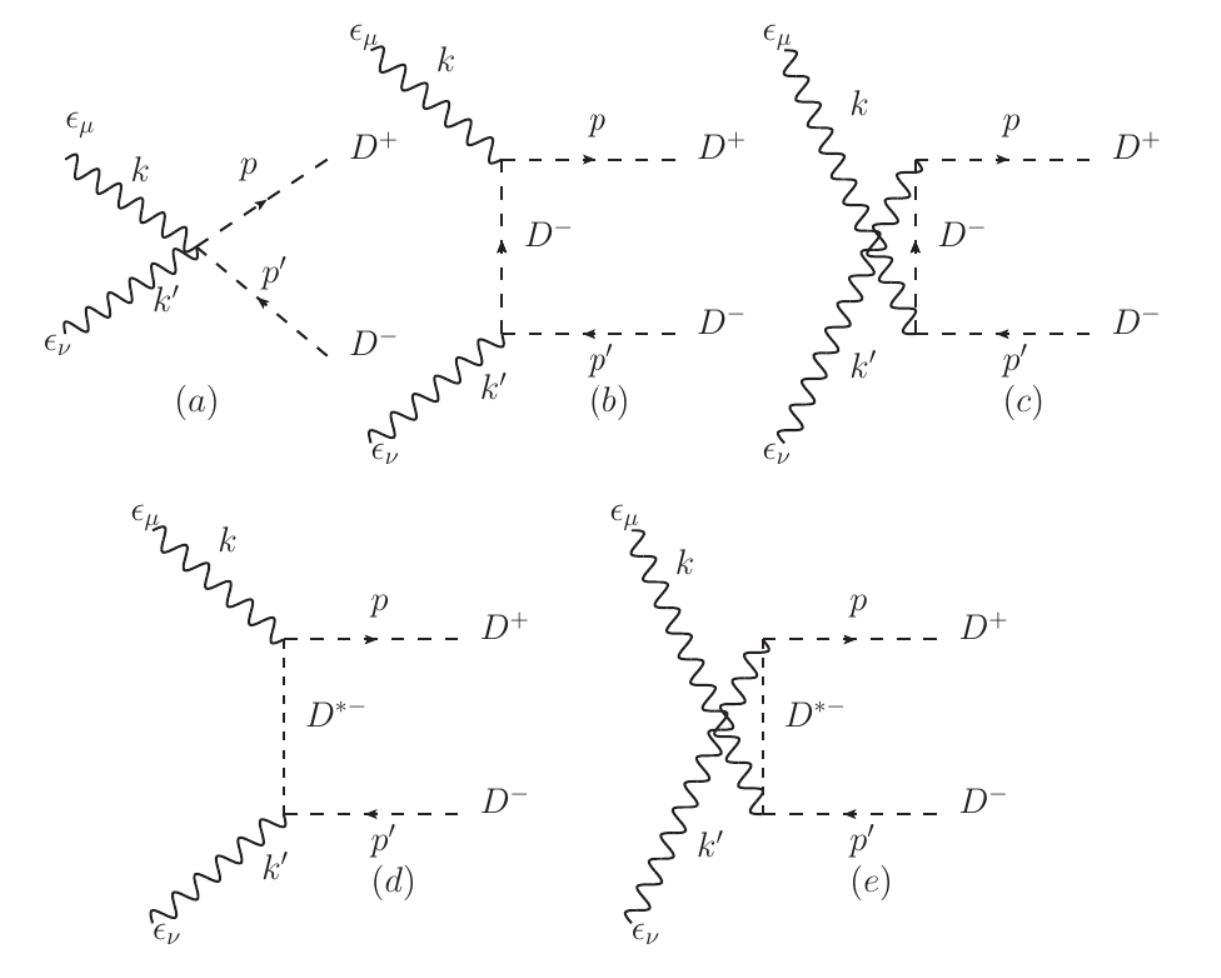}
\caption{Feynman diagrams for the process $\gamma\gamma\rightarrow D^+D^-$.}
    \label{fig:diagramas}
\end{figure}

While the relation between these two mechanisms (and whether they are       
complementary or equivalent) remains to be explored, we choose to work with 
the hadronic formalism. Along this line, we will study the process 
$\gamma \gamma \to D^+ D ^-$ with the Lagrangian densities \cite{lag}
\begin{equation}
    \mathcal{L} = (D_\mu\phi)^*(D^\mu\phi) - m_D^2\phi^*\phi - 
\frac{1}{4}F_{\mu\nu}F^{\mu\nu} \, ,
\end{equation}
and 
\begin{equation}
    \mathcal{L} = -ig_{\gamma D^{+}D^{*-}} F_{\mu\nu}\epsilon^{\mu\nu\alpha\beta}(D^{*-}_\alpha\overset{\leftrightarrow}{\partial_\beta}D^{+} + D^-\overset{\leftrightarrow}{\partial_\beta}D^{*+}_{\alpha}) \, ,
\end{equation}
where
\begin{equation}
    D_\mu\phi = \partial_\mu\phi + ieA_\mu\phi \,\, ,    \hskip1cm   
    F_{\mu\nu} = \partial_\mu A_\nu - \partial_\nu A_\mu \nonumber \, ,
\end{equation}
and $\phi$, $D^*$ and $A_{\mu}$ represent the $D^+$ (or $D^-$), 
the $D^{*+}$ (or $D^{*-}$)    and the 
photon fields, respectively. The Feynman rules can be derived from the 
interaction terms and they yield the Feynman diagrams for the process 
$\gamma \gamma \to D^+ D^-$ shown in Fig. \ref{fig:diagramas}.  In the figure 
we also show the quadrimomenta of the incoming photons 
$k^{\mu} = (E_{p},0,0,k)$, ${k'}^{\mu} = (E_{k'},0,0, k')$ 
and of the outgoing mesons $p^{\mu} = (E_p,0,0,p)$, 
${p'}^{\mu} = (E_{p'},0,0,p')$. 
The total amplitude is  given by:
\begin{equation}
  iM =  iM_{(a)} +  iM_{(b)} +  iM_{(c)}   iM_{(d)} +  iM_{(e)} \, ,
\label{tamp}
\end{equation}
where
\begin{align}
iM_{(a)} &= 2ie^2g_{\mu\nu} F(\Bar{q}^2)F(\Bar{q}^2)
            \varepsilon^{*\mu}(k) \varepsilon^{*\nu}(k') \, ,\\
iM_{(b)} &= \varepsilon^{*\mu}(k)ie F(\hat{t}) (-2p_\mu + k_\mu)
\frac{i}{(k-p)^2-m_D^2}ie F(\hat{t})(2p'_\nu - k'_\nu)
\varepsilon^{*\nu}(k') \, ,\\
iM_{(c)} &= \varepsilon^{*\mu}(k')ieF(\hat{u})(-2p_\mu + k'_\mu)
\frac{i}{(k'-p)^2-m_D^2}ieF(\hat{u})(2p'_\nu - k_\nu)
\varepsilon^{*\nu}(k) \, , \\
iM_{(d)} &= \varepsilon^{*}_{\mu}(k)[-2g\epsilon^{\sigma\mu\alpha\rho}k_{\sigma}(k_\rho - 2p_\rho)F(\hat{t})]\left[\frac{-i(g_{\alpha\beta} - \frac{(k-p)_\alpha (k-p)_\beta}{m_{D^*}^2})}{(k-p)^2-m_{D^*}^2}\right][2g\epsilon^{\delta\nu\beta\lambda}k'_{\delta}(-k'_\lambda + 2p'_\lambda)F(\hat{t})]\varepsilon^{*}_{\nu}(k')\, , \\
iM_{(e)} &= \varepsilon^{*}_{\mu}(k')[-2g\epsilon^{\sigma\mu\alpha\rho}k'_{\sigma}(k'_\rho - 2p_\rho)F(\hat{u})]\left[\frac{-i(g_{\alpha\beta} - \frac{(k'-p)_\alpha (k'-p)_\beta}{m_{D^*}^2})}{(k'-p)^2-m_{D^*}^2}\right][2g\epsilon^{\delta\nu\beta\lambda}k_{\delta}(-k_\lambda + 2p'_\lambda)F(\hat{u})]\varepsilon^{*}_{\nu}(k)\, , 
 \end{align}
where $\Bar{q}^2 = [(k-p)^2+(k'-p)^2]/2$ and 
$g = g_{\gamma D^{+}D^{*-}}=-0.035$ \cite{lag}.
We have introduced the Mandelstam 
variables of the elementary process, which are $ \hat{s}=(k + k')^2$, 
$\hat{t} = (k - p)^2$ and $\hat{u} = (k' - p)^2$.  As usual, we
have included form factors, $F(q)$, in the vertices of the above amplitudes. 
We shall follow \cite{kk} and use the monopole form factor given by
\begin{equation}
     F(q^2) = \frac{\Lambda^2 - m^2_{D^{(*)}} }{\Lambda^2 - q^2} \, , 
\end{equation}
where $q$ is the 4-momentum of the exchanged meson and $\Lambda$ is a cut-off 
parameter.  This choice has the advantage of yielding 
automatically $F(m^2_D) = 1$ and $F(m^2_{D^*}) = 1$ when the exchanged meson 
is on-shell. 
The above form is arbitrary but there is 
hope to improve this ingredient of the calculation using QCD sum rules to 
calculate the form factor, as done in \cite{ff}, thereby reducing the 
uncertainties. Taking the square of the amplitude Eq. (\ref{tamp}) 
and the average over the photon polarizations 
it is straigthforward to calculate the differential cross section:
\begin{equation}
\frac{d\sigma}{d\Omega} = \frac{1}{64\pi^2}\frac{1}{E_{CM}^2}
\frac{|{\bf p}|}{|{\bf k}|}|M(\gamma\gamma\rightarrow D^+D^-)|^2 \, ,
\label{dsig}
\end{equation}
In the center-of-mass reference frame we have
${\bf k}=-{\bf k'}$ and hence ${\bf p} = -{\bf p'}$,  
$E_{CM} = E_k + E_{k'} = 2|{\bf k}|$ and 
$E_{CM}= E_p + E_{p'} = 2\sqrt{|{\bf p}|^2+m_{D}^2} $. 
It is then easy to see that: 
\begin{equation}
\frac{|{\bf p}|}{|{\bf k}|} = \sqrt{\frac{(E_{CM}^2-4m_D^2)/4}{E_{CM}^2/4}}
= \sqrt{1 - \frac{4m_D^2}{E_{CM}^2}} \, ,
\label{pok}
\end{equation} 
Inserting Eq. (\ref{pok}) into Eq. (\ref{dsig}) and using $E^2_{CM} = \hat{s}$  
we find: 
\begin{equation}
\label{crossfp}
\sigma = \frac{1}{64\pi^2}\frac{1}{\hat{s}}\sqrt{1 - \frac{4m_D^2}{\hat{s}}}
\int |M(\gamma \gamma\rightarrow D^+D^-)|^2d\Omega\, .
 \end{equation}
The angular integral can be done using the relations:
$$
\hat{t}= m_D^2 - \frac{\hat{s}}{2} + \left(\sqrt{\hat{s}(\frac{\hat{s}}{4} - m_D^2)}\right)\cos(\theta),\quad 
\hat{u}= m_D^2 - \frac{\hat{s}}{2} - \left(\sqrt{\hat{s}(\frac{\hat{s}}{4} - m_D^2)}\right)\cos(\theta),
$$
where $\theta$ is the angle between ${\bf k}$ and  ${\bf p}$.
We emphasize that the only unknown in our calculation is the cut-off 
parameter $\Lambda$. In what follows, we will determine it  fitting our cross
section to the LEP data on the process $e^+ e^- \to e^+ e^- c \bar{c}$. 

\section{Production of bound states}

Now we describe the method to construct a bound state (denoted $B$) from
the $D^+ D^-$ pair. As in \cite{pp1}, we impose phase space constraints on
the mesons, forcing them to be ``close together''. 
Here we do this through the 
prescription discussed in \cite{pes}. The bound state $|B \rangle$ is 
defined as
\begin{equation}
\frac{|B\rangle}{\sqrt{2E_B}}\equiv \int
\frac{d^3q}{(2\pi)^3}\Tilde{\psi^*}({\bf q})
\frac{1}{\sqrt{2E_q}}\frac{1}{\sqrt{2E_{-q}}}|{{\bf q},-{\bf q}\rangle },
\label{bound}
\end{equation}
where $E_B$ is the bound state energy, ${\bf q}$ is the relative three momentum
between  $D^+$ and $D^-$ in the state $B$, $E_{\pm q}$ are the energies of
$D^+$ and $D^-$ and $\Tilde{\psi}({\bf q})$ is the bound state wave function
in momentum space, which has the following properties:
\begin{equation}
\Tilde{\psi}({\bf q}) = \int d^3xe^{i{\bf q}\cdot{\bf x}}\psi({\bf x}); 
\hskip1.0cm
\int \frac{d^3q}{(2\pi)^3}|\Tilde{\psi}({\bf q})|^2 = 1.
\end{equation}  
From Eq. (\ref{bound}), we can write the following relation between the 
amplitudes:
\begin{equation}
\frac{M(\gamma\gamma\rightarrow B)}{\sqrt{2E_B}}= \int \frac{d^3q}{(2\pi)^3}
\Tilde{\psi}^*({\bf q})\frac{1}{\sqrt{2E_{D^+}}}\frac{1}{\sqrt{2E_{D^-}}}
M(\gamma\gamma\rightarrow D^+D^-),
\end{equation}
We assume that the  ${\bf p} \simeq {\bf p'}$ and hence 
$E_{D^+} \simeq E_{D^-} = E_D$ and also 
${\bf q} = {\bf p} - {\bf p'} \simeq 0 $. 
Therefore the energy $E_D$ and the amplitude 
$M(\gamma\gamma\rightarrow D^+D^-)$ can be taken out of the integral. 
Moreover, 
since the binding energy is small we have $E_B \simeq 2E_D$ and hence
\begin{align}
\frac{M(\gamma\gamma\rightarrow B)}{\sqrt{2E_B}}&= \frac{M(\gamma\gamma
\rightarrow D^+D^-)}{E_B}\int \frac{d^3q}{(2\pi)^3}\Tilde{\psi}^*
({\bf q}),\nonumber\\
M(\gamma\gamma\rightarrow B) &= \sqrt{\frac{2}{E_B}}
M(\gamma\gamma\rightarrow  D^+D^-)\int \frac{d^3q}{(2\pi)^3}
\int d^3x \, \psi^* ({\bf x})e^{i{\bf q}\cdot{\bf x}},\nonumber\\
&= \sqrt{\frac{2}{E_B}}M(\gamma\gamma\rightarrow D^+D^-)\int d^3x \, \psi^*
({\bf x}) \, \delta^{(3)}({\bf x}) , \nonumber \\
&= \psi^*(0)\sqrt{\frac{2}{E_B}}M(\gamma\gamma\rightarrow D^+D^-) \, .
\label{project} 
\end{align}
With the amplitude above we calculate the cross section for bound state
production:  
\begin{equation}
d\sigma = \frac{1}{H}\frac{d^3p_B}{(2\pi)^3}\frac{1}{2E_B}(2\pi)^4
\delta^{(4)}(k+k'-p_B)|M(\gamma\gamma\rightarrow B)|^2,
\label{sigbs}
\end{equation}
where $p_B$ is the momentum  of the produced bound 
state and $H$ is the flux factor. 
Now we will work in the 
center of mass  frame of the $AA \to AAB$ collision, in which 
the momenta of the incoming photons may be different. In this frame we have
\begin{equation}
k = (\omega_1,0,0,\omega_1) \, , \hskip1.0cm
k' = (\omega_2,0,0,-\omega_2) \, ,  \hskip1.0cm
p_B \equiv p+p' = (E_B,0,0,\omega_1-\omega_2) \, , 
\end{equation}
where $E_B = \sqrt{(\omega_1-\omega_2)^2 + m_B^2}$ and $\omega_1$ and 
$\omega_2$ are the energies of the colliding photons. 
The flux factor is then 
given by
\begin{equation}
H = 4\sqrt{(k\cdot k')^2 - m_k^2m_{k'}^2} = 4 k\cdot k' 
= 4 (k_0k'_0 - {\bf k}\cdot{\bf k'})=4 (\omega_1\omega_2 - \omega_1(-\omega_2)) 
= 2(4\omega_1\omega_2) \, ,
\end{equation}
Inserting this expression into Eq.(\ref{sigbs}) and integrating, 
the cross section reads
\begin{align}
\sigma(\omega_1,\omega_2) &= \frac{2\pi}{2(4\omega_1\omega_2)}\int
\frac{d^3p_B}{2E_B}\delta(E_{CM} - E_B)\delta^{(3)}({\bf k}+{\bf k'}-
{\bf p}_B)\left[\frac{2}{E_B}|\psi(0)|^2|M(\gamma\gamma\rightarrow D^+D^-)
|^2\right], \nonumber\\
&= \frac{\pi |\psi(0)|^2}{4\omega_1\omega_2E_B^2}
|M(\gamma\gamma\rightarrow D^+D^-)|^2 \frac{4 \omega_1^2 + m_B^2}{8 \omega_1^2}
\delta(\omega_2 - \frac{m_B^2}{4 \omega_1})  \, ,   
\label{sigmabound2}
\end{align}
where we have used that $E^2_{CM} = 4 \omega_1 \omega_2$. 

To proceed with the calculation we need to know the bound state wave 
function at the 
origin $|\psi (0)|^2$.  
Fortunately, in \cite{go} a similar bound state of open charm mesons was  
studied with the Bethe-Salpeter equation and an expression for the wave 
function was derived. In the first part of their paper the authors present 
a formalism which  is general and can be adapted to our system. Formally, 
the Bethe-Salpeter equation reads $T = V + VGT$, where $T$ is the two-body 
amplitude, $V$ is a matrix with elements $V_{ij}$ which are the amplitudes of 
the $i \to j$ transitions and which are calculated from a given effective 
Lagrangian. Finally $G$ is a loop function, which can be regularized with a 
cut-off. Here we will just quote the main formulas needed to 
calculate $\psi(0)$, which is given by
\be
\psi(0) = \frac{g}{(2\pi)^{3/2}} G \, ,
\label{psi0}
\ee
where
\be
G = -8\mu\pi\left(\Lambda_0 - \gamma \arctan\left(\frac{\Lambda_0}
{\gamma}\right)\right) \, , 
\hskip1cm \gamma = \sqrt{2\mu E_{b}} \, , \hskip1cm  
g^2= \frac{\gamma}{8\pi\mu^2(\arctan(\frac{\Lambda_0}{\gamma}) -
\frac{\gamma \Lambda_0}{\gamma^2 + \Lambda_0^2})}.
\label{gs}
\ee
In the above expressions $\mu$ is the reduced mass ($\mu = m_D /2$),  
$\Lambda_0$ is a cut-off parameter and $E_{b}$ is the binding energy.
We shall follow \cite{mb} and assume that $\Lambda_0 = 1$ GeV. From the 
above equations we see that one can compute the (dynamically generated) mass 
of a bound state and then determine its binding energy. Knowing $\mu$, $E_b$ 
and fixing $\Lambda_0$, we can use the above formulas to calculate $\psi(0)$.  
In what follows our reference value will be obtained using  $m_D =1870$ MeV 
and the mass of the bound state equal to  $M_B = 3723$ MeV, as  found in 
\cite{mb}. With these numbers we get  
$E_{b}=17$ MeV and $|\psi(0)|^2 = 0.008$ GeV$^3$. These will be the values
used to obtain all results, unless stated otherwise.

\section{Equivalent photon approximation and the number of photons}

The equivalent photon approximation is well known and it is described in 
several papers \cite{epa,epa2}. In general, when the photon source is a nucleus
one has to use form factors and the calculation becomes somewhat complicated.
Here we will follow \cite{epa} and define an UPC in momentum space. 
The distribution of equivalent photons generated by a moving particle with 
the charge $Ze$ is \cite{epa}: 
\beq
n({\bf q}) d^3 q = \frac{Z^2 \alpha}{\pi^2} \frac{({\bf q}_{\perp})^2}
{\omega \, q^4} d^3 q = \frac{Z^2 \alpha}{\pi^2 \omega} 
\frac{({\bf q}_{\perp})^2}{\left( ({\bf q}_{\perp})^2 
+ (\omega/\gamma)^2\right)^2} d^3 q  \, , 
\label{defn}
\eeq
where $\alpha= e^2/(4 \, \pi)$,  $q$ is the photon 4-momentum, 
${\bf q}_{\perp}$ is its transverse 
component, $\omega$ is the photon energy and $\gamma$ is the Lorentz factor 
of the photon source ($\gamma = \sqrt{s}/2m_p$ and $m_p$ is the proton mass).
To obtain the equivalent photon spectrum, one has to integrate this expression 
over the transverse momentum up to some value $\hat{q}$. The value of $\hat{q}$
is given by $\hat{q} = \hbar c/ 2R$, where $R$ is the radius of the projetile.
For Pb, $R \approx 7$ fm and hence $\hat{q} \approx 0.014$ GeV. After the 
integration over the photon transverse momentum the equivalent photon energy 
spectrum  is given by:
\begin{equation}
n(\omega)d\omega = \frac{2Z^2\alpha}{\pi}\ln\left(\frac{\hat{q}\gamma}{\omega}
\right)\frac{d\omega}{\omega},
\label{nomega}
\end{equation}
Because of the approximations  the above distribution is valid when the 
condition  $\omega \ll \hat{q}\gamma$
is fullfiled. Using Eq. (\ref{nomega}) we can compute  the cross sections of
free pair production, $\sigma_P$, and of bound state production, $\sigma_B$. 
They are given by: 
\begin{align}
\sigma_P(A \, A \rightarrow A \, A \,  D^+D^-) 
&= \int\limits_{m_D^2/\hat{q}\gamma}^
{\hat{q}\gamma}  d\omega_1 \int\limits_{m_D^2/\omega_1}^{\hat{q}\gamma}
d\omega_2 \, \sigma_{P}(\omega_1,\omega_2) \, n(\omega_1) \, n(\omega_2),
\label{sigmafp2}\\
\sigma_B (A \, A \rightarrow A \, A \,  B) &= \int\limits_{m_D^2/\hat{q}\gamma}^
{\hat{q}\gamma}d\omega_1\int\limits_{m_D^2/\omega_1}^{\hat{q}\gamma}
d\omega_2 \, \sigma_{B}(\omega_1,\omega_2) \, n(\omega_1) \, n(\omega_2),
\label{sigmabs2}
\end{align}
where $\sigma_{P}(\omega_1,\omega_2)$ and $\sigma_{B}(\omega_1,\omega_2)$ are 
given by Eqs. (\ref{crossfp}) (with $\hat{s} = 4 \omega_1 \omega_2$)  
and (\ref{sigmabound2}) respectively.

\section{The low energy approximation}

\subsection{Free pairs}

At low photon energies and close to the $D^+ D^-$ threshold, the produced 
mesons are non-relativistic and we can use the approximation 
$k - p \approx (0, 0, 0, m_D)$ in the heavy meson propagator, i.e.:
$$
\frac{1}{(k-p)^2 - m_D^2} \approx \frac{1}{0-m_D^2-m_D^2} 
= \frac{-1}{2m_D^2} \, .
$$
An analogous expression can be written for the $D^*$ propagator. 
From the above relation we can see that in this low energy regime the 
amplitudes with propagators are proportional to $\propto 1/m_D^2$ 
(Figs.~\ref{fig:diagramas}b and  \ref{fig:diagramas}c)
and $\propto 1/m_D^{*2}$ (Figs.~\ref{fig:diagramas}d and \ref{fig:diagramas}e)
and 
can be neglected when compared to the amplitudes without propagators, 
such as the one of the contact interaction in Fig.~\ref{fig:diagramas}a. 
With this approximation 
the amplitude for  $D^+D^-$ production in the  process 
$\gamma \gamma\rightarrow D^+D^-$  is given by:
\begin{equation}
iM(\gamma\gamma\rightarrow D^+D^-) \approx 2ie^2 F^2(-m_D^2) g_{\mu\nu}
\varepsilon^{*\mu}(k)\varepsilon^{*\nu}(k') \, .
\label{ampfp}\\
\end{equation}
Taking the square and performing the average over the photon polarizations
we have:
\begin{align}
|M(\gamma\gamma\rightarrow D^+D^-)|^2 &= \frac{1}{4}\sum_{pol}
2ie^2 F^2(-m_D^2)g_{\mu\nu}\varepsilon^{* \mu}(k)
\varepsilon^{*\nu}(k') \,\, 
(-2ie^2) F^2(-m_D^2) g_{\sigma\rho}
\varepsilon^{\sigma}(k)\varepsilon^{\rho}(k') \, ,
\nonumber\\
&= e^4 F^4(-m_D^2) g_{\mu\nu}g_{\sigma\rho}g^{\mu\sigma}g^{\nu\rho}
= 4e^4 F^4(-m_D^2) \, .
\label{ampfp2}
\end{align}
Inserting this amplitude into Eq. (\ref{dsig})
we find: 
\begin{equation}
\frac{d\sigma}{d\Omega} = \frac{e^4 F^4(-m_D^2)}{16\pi}
\frac{1}{E_{CM}^2}\sqrt{1 - \frac{4m_D^2}{E_{CM}^2}}.
 \end{equation}
Performing the integral over the solid angle and using the definitions
$\alpha \equiv e^2/4\pi$ and $E_{CM}^2 = 4\omega_1\omega_2$, we find
\begin{equation}
\label{sigmafp}
\sigma (\omega_1,\omega_2) = \frac{\pi\alpha^2 F^4(-m_D^2)}{\omega_1\omega_2} 
\sqrt{1 - \frac{m_D^2}{\omega_1\omega_2}} \, , 
\end{equation}
which is then substituted in Eq. (\ref{sigmafp2}) to give the final cross 
section for $A \, A \rightarrow A \, A \,  D^+D^-$. 

\subsection{Bound states} 

In the low energy approximation the produced bound state is non-relativistic
and then Eq. (\ref{project}) reduces to:
\begin{equation}
M(\gamma\gamma\rightarrow B)
= \psi^*(0)\sqrt{\frac{2}{m_B}}M(\gamma\gamma\rightarrow D^+D^-) \, . 
\label{proj}
\end{equation}
Inserting Eq. (\ref{ampfp2}) into the above equation and then using it in 
Eq. (\ref{sigmabound2}) we have: 
\begin{equation}
\sigma(\omega_1,\omega_2) = 
\frac{32\pi^3 \alpha^2 F^4(-m_D^2) |\psi(0)|^2}{m_B}\frac{1}{\omega_1\omega_2}
\delta(4\omega_1\omega_2 - E_B^2).
\label{sigmabound}
 \end{equation}
Substituting the above expression into Eq. (\ref{sigmabs2}) and integrating 
we obtain the final analytical expression:
\begin{equation}
\sigma_B (A \, A \rightarrow A \, A \,  B) = \frac{256\pi 
|\psi(0)|^2 Z^4 \alpha^4 F^4(-m_D^2)}
{3m_B^5}\left[\ln\left(\frac{s \hat{q}^2}{ m_p^2 m_B^2}\right)\right]^3 \, .  
\label{anal}
\end{equation}
We emphasize that ``low energy'' here refers to the energy released by the 
projectiles, i.e. the invariant mass of the photon pair. The nuclear 
projectiles themselves may have very high energies. 

\begin{figure}
\centering
\includegraphics[width=.60\linewidth]{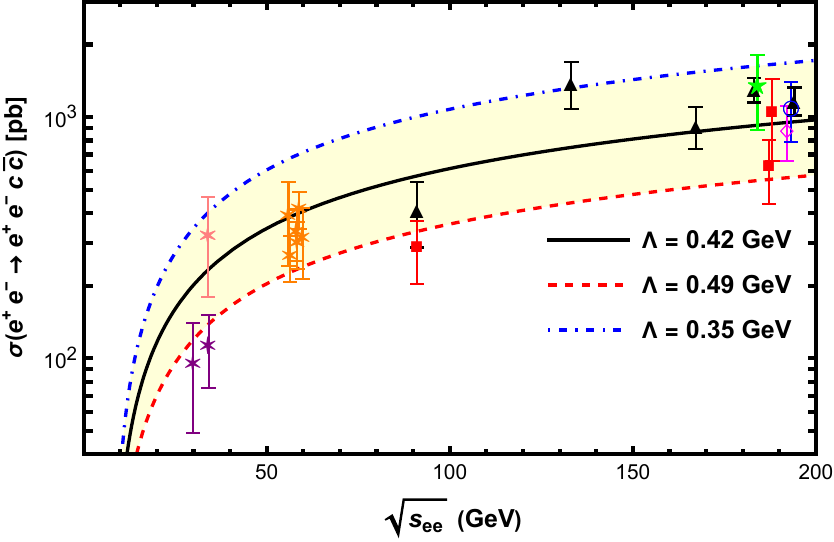}
\caption{Cross section for  the process $e^+ e^- \to c \bar{c}$ as a 
function of the energy $\sqrt{s}$ measured by the LEP Collaborations. 
Data are from Refs. \cite{lep}. Purple stars from TASSO, pale red single
star from JADE, bright orange stars from TOPAZ, AMY and VENUS, triangles
from L3, squares and the green star from ALEPH, single diamond from DELPHI
and circle from OPAL. The curves are calculated with Eqs. (\ref{crossfp})
and (\ref{sigmafp2}).}
\label{figdat}
\end{figure}
\begin{figure}
\begin{tabular}{cc}
\includegraphics[width=.45\linewidth]{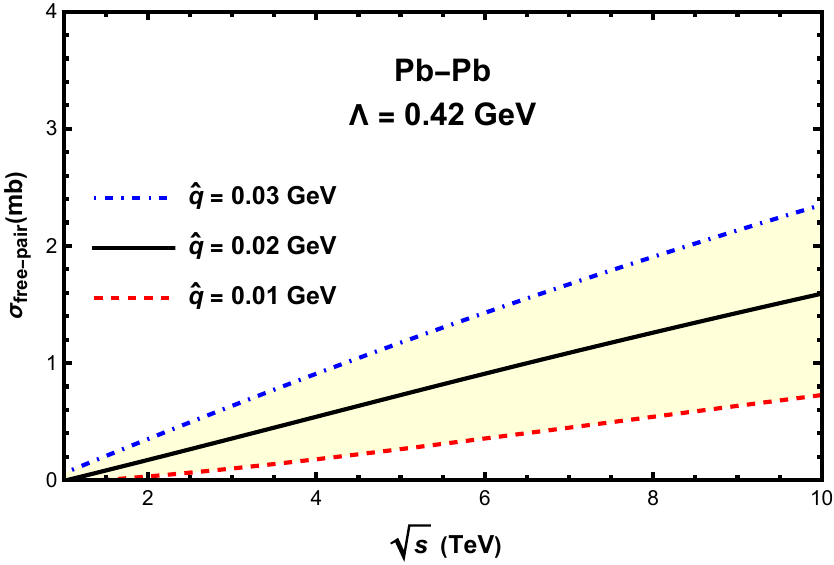}&
\includegraphics[width=.45\linewidth]{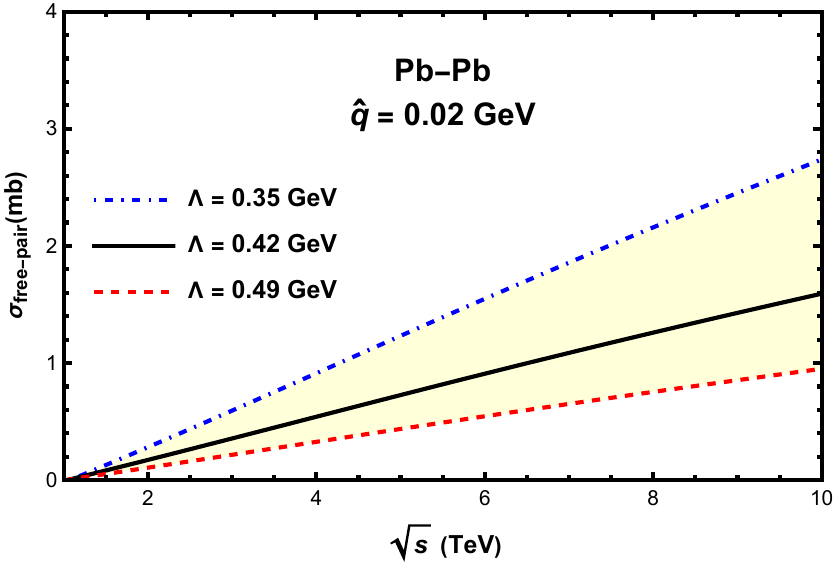} \\
  (a) & (b)
\end{tabular}
\caption{Cross sections for free $D^+ D^-$ pair production as a function of the 
energy $\sqrt{s}$. a) Dependence on $\hat{q}$ for fixed $\Lambda$. 
b) Dependence on $\Lambda$ for fixed $\hat{q}$.} 
\label{sigmal}
\end{figure}
\begin{figure}
\begin{tabular}{ccc}
\includegraphics[width=.33\linewidth]{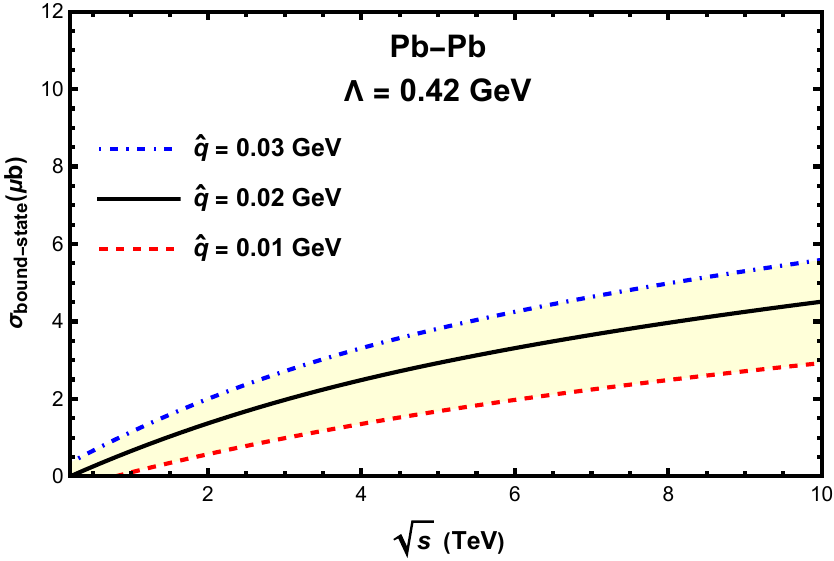}&
\includegraphics[width=.33\linewidth]{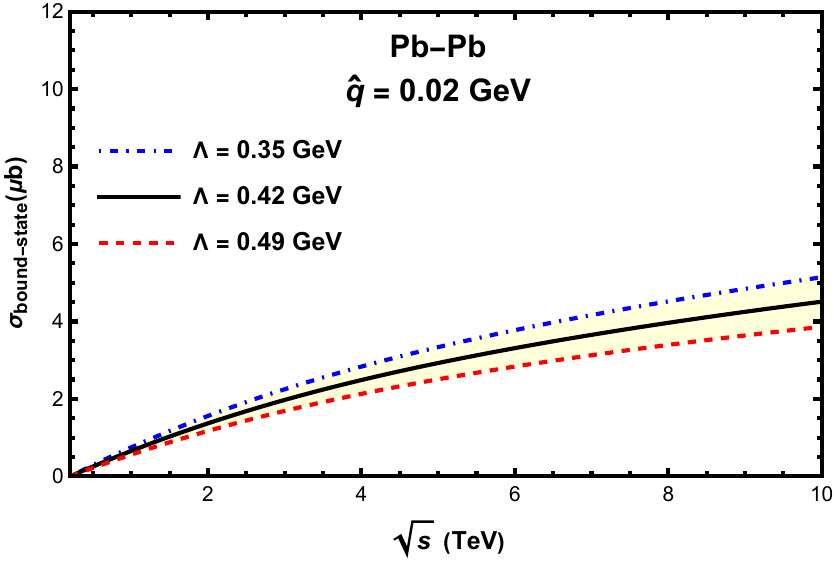}&
\includegraphics[width=.33\linewidth]{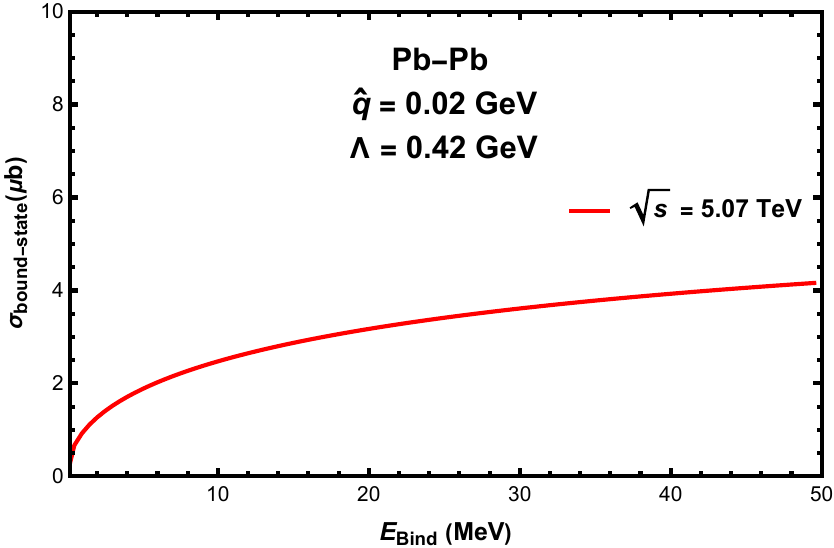}\\ 
(a) & (b) & (c) 
\end{tabular}
\caption{Cross sections for $D^+ D^-$ bound state production as a function 
of the energy $\sqrt{s}$. a) Dependence on $\hat{q}$ for fixed $\Lambda$.
b) Dependence on $\Lambda$ for fixed $\hat{q}$.
c) Dependence on the binding energy for fixed $\Lambda$ and $\hat{q}$. 
}
\label{sigmab}
\end{figure}
\begin{figure}
\begin{tabular}{ccc}
\includegraphics[width=.33\linewidth]{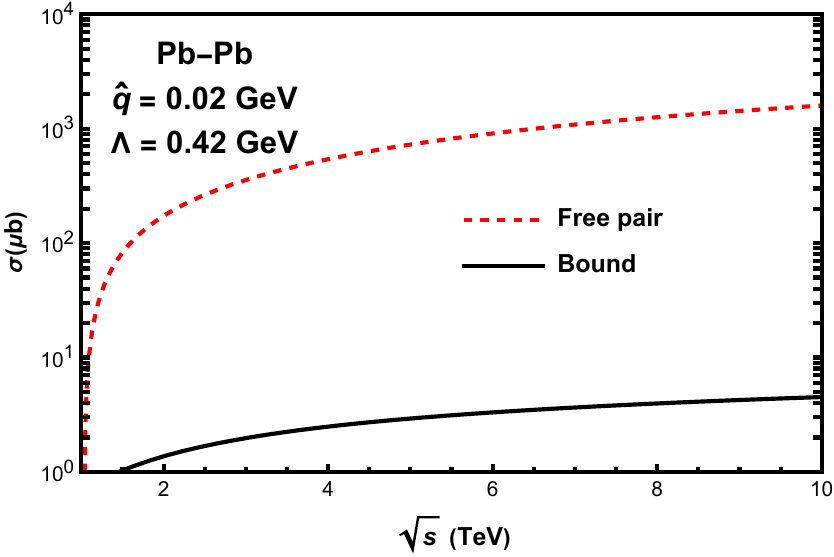}&
\includegraphics[width=.33\linewidth]{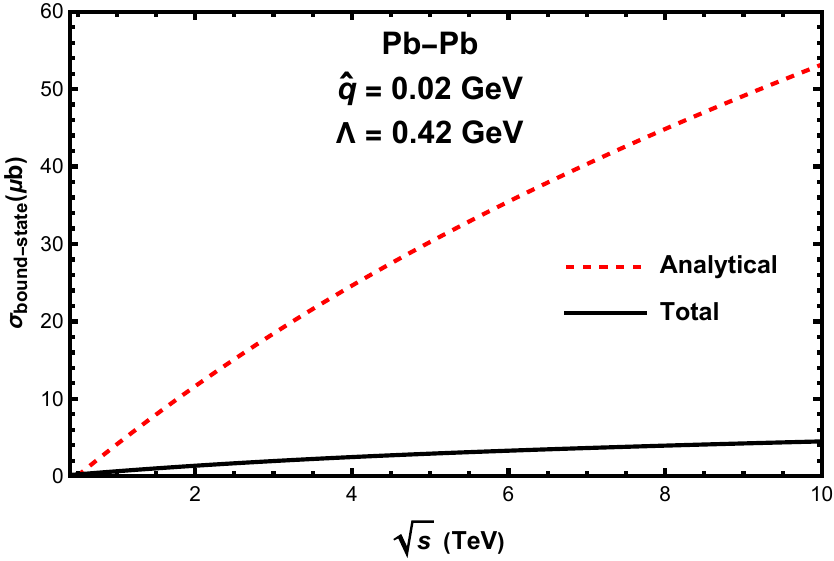}&
\includegraphics[width=.33\linewidth]{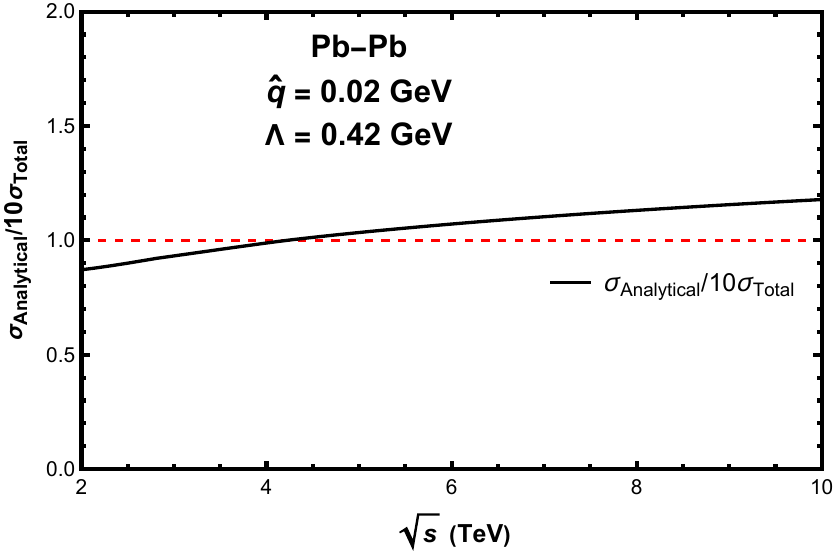}\\
(a) & (b) & (c)
\end{tabular}
\caption{Cross sections as a function of  $\sqrt{s}$. a) Comparison between 
the free $D^+D^-$ pair and  bound state $B$ cross sections. b) Comparison 
between the complete numerical solution and the approximate analytical cross 
section. c) Ratio between the cross section.}
\label{sigmas}
\end{figure}

\section{Numerical results and discussion} 

Having derived all the main formulas and discussed the numerical inputs, now
we present our numerical results. In Fig.~\ref{figdat} we show the cross 
sections for free pair production and compare it to the existing experimental 
data from LEP \cite{lep}. In fact, the LEP data are for 
$e^+ \, e^- \to e^+ \, e^- \, c \, \bar{c}$, i.e., the measured final states are
$D^+ D^-$ and $D^0 \bar{D}^0$.  We assume that these two final states have the
same cross section and, in order to compare with the data, we multiply our
cross section $ \sigma( e^+ \, e^- \to e^+ \, e^- \, D^+ \, D^-) $ by a factor 
two. In order to fit these data we
will adapt expression (\ref{sigmafp2}) to  electron-positron collisions. The   
$\gamma \gamma \to D^+ D^-$ cross section is same but the  photon flux 
from the electron (and also from the positron) and the integration limits are  
different. The adaptation of Eq. (\ref{sigmafp2}) is performed in Appendix 
\ref{apa}.  
Comparing our formula with  
these data, we determine the only parameter in the calculation, which is the 
cut-off $\Lambda$. In the figure, the curves are obtained  substituting 
Eqs. (\ref{crossfp}) and (\ref{newn}) into (\ref{sigmafp2}). In the latter 
$\hat{q} = m_e$. We did not attempt to perform a least chi square fit. Instead
we will carry on some uncertainty and work with the band 
$0.35 < \Lambda <0.49$ GeV.

In Fig. \ref{sigmal} we show the cross section for $D^+ D^-$ production.
The black solid lines show the result with our central parameter choice. 
Fig. \ref{sigmal}a shows the sensitivity of the result to the value of
$\hat{q}$. In Fig. \ref{sigmal}b we vary the values of $\Lambda$ in the range
defined in Fig.~\ref{figdat}. In this sense we propagate the uncertainty 
from the fit of the data to our results. Taking this as the error in our 
result, the obtained cross section for the reaction 
$Pb \, Pb \rightarrow Pb \, Pb \, D^+ D^-$ at $\sqrt{s_{NN}} = 5.02$ TeV 
is:
\begin{equation}
\sigma (Pb \, Pb \rightarrow Pb \, Pb \, D^+ D^-) = 
0.75^{+0.4}_{-0.4} \,\, \mbox{mb}. 
\label{sigpfinal}
\end{equation}
Assuming that the reaction $Pb \, Pb \rightarrow Pb \, Pb \, D^0 \bar{D}^0$
has the same cross section as the one given above for charged states, the 
total cross section for charm production in photon-photon exclusive processes, 
we have
\begin{equation}
\sigma_{exclusive} (Pb \, Pb \rightarrow Pb \, Pb \, c \bar{c}) =
1.5^{+0.4}_{-0.4} \,\, \mbox{mb}. 
\label{sigccex}
\end{equation}
In  Appendix \ref{apb}, using crude approximations, we have arrived at the 
following identity for inclusive cross sections: 
\begin{equation}
\sigma_{QED}(Pb \, Pb \rightarrow Pb \, Pb \, c \, \bar{c} \, X)
\approx \sigma_{QCD}(p \, p \rightarrow p \, p \, c \, \bar{c} \, X) \, .
\label{estima}
\end{equation} 
This suggests that, for a given high energy, electromagnetic interactions 
in Pb-Pb are as efficient as strong interactions in p-p for charm production.
Very recently, two independent analyses of the LHC data obtained
estimates for the inclusive charm production cross section in 
proton proton collisions at $\sqrt{s} = 5.02$ TeV, which are dominated by the 
strong interaction. In \cite{yg} the authors 
find:
\begin{equation}
 \sigma_{inclusive}(p \, p \rightarrow  c \, \bar{c} \, X)
= 8.43^{+1.05}_{-1.16} \,\,  \mbox{mb}.
\label{yang}
\end{equation}
and a quite similar result was obtained in \cite{lu}.  
According to (\ref{estima}), (\ref{sigccex}) should be smaller (because it is
exclusive) but of the order of 
magnitude of (\ref{yang}). Considering the uncertainties, 
we believe that this is approximately true.


In Fig.~\ref{sigmab} we present the cross section for bound state 
production and study its dependence on $\hat{q}$ (Fig.~\ref{sigmab}a), on  
$\Lambda$  (Fig.~\ref{sigmab}b) and on the binding energy $E_b$ 
(Fig.~\ref{sigmab}c).  As expected, it is much 
smaller than the cross section for open free pair production. However, 
it is encouraging to see that at $\sqrt{s_{NN}} \approx 5.02$ TeV we have:
\begin{equation}
\sigma (Pb \, Pb \rightarrow Pb \, Pb \, B) =
3.0^{+0.8}_{-1.2} \,\, \mu \mbox{b} \, . 
\label{sigbfinal}
\end{equation}
This number should be compared with results found in \cite{br} and in 
\cite{fa}. In those papers, the production cross section of scalar 
states $X(3940)$ and $X(3915)$ in Pb Pb  ultra-peripheral collisions at 
$\sqrt{s_{NN}} = 5.02$ TeV were calculated and the results were in the range
\begin{equation}
5 \leq \sigma (Pb \, Pb \rightarrow Pb \, Pb \, R) \leq 11 \,\, \mu b \, . 
\label{sigx}
\end{equation}
where $R$ stands for $X(3940)$ or $X(3915)$. The works \cite{br} and \cite{fa}
are relevant for us because there the states $R$ were also treated as molecules. 
However there is an important difference. In \cite{br} and \cite{fa} the 
authors used the Low formula, which connects the $\gamma \gamma \to R$ cross
section with the $R \to \gamma \gamma$ decay width (used as input). 
In very few cases this width 
was measured and in some other very few cases the width was estimated with the 
help of a formalism valid for dynamically generated (and hence molecular) 
states. Here we propose a method to form the molecular state which is  more 
general and independent of the knowledge of the decay width. Another difference 
is 
that the states $X(3940)$ and $X(3915)$ are significantly heavier than the 
$D^+ D^-$ molecule, whose mass is $3723$ MeV. Moreover, in \cite{br} and 
\cite{fa} the equivalent photon calculation was done in the impact parameter 
space. In spite of these differences the obtained cross sections are of the 
same order of magnitude.

For completeness, in  Fig.~\ref{sigmas}a we compare the cross sections for 
free pair and bound state production and  in  Fig.~\ref{sigmas}b we compare the 
exact numerical evaluation of $\sigma_B$ with the approximate analytical     
expression, Eq.(\ref{anal}). We observe that the cross section obtained with 
the analytical formula is  accurate only at low energies. At higher 
energies it becomes larger than the complete numerical formula. We can 
understand this behavior  noticing that in Eq.(\ref{anal}) we assumed that 
both $E_B$ and the form factor $F(q^2)$ did not depend on $\omega_1$ nor on 
$\omega_2$ at low energies and therefore resulted in a smaller denominator 
($m_B$ where it should have been $E_B$) and a constant argument of the form 
factor, which, as we can see from Fig.~\ref{sigmab}b, is crucial to our 
numerical results. Nevertheless, the exact and the analytical formula differ
essentially only by a multiplicative factor close to 10. Dividing 
Eq. (\ref{anal}) by 10, it reproduces the exact formula within ~ 20 \% 
accuracy in the 
relevant LHC range and can thus be useful for practical applications. This is 
shown Fig.~\ref{sigmas}c.

To summarize, we have calculated, for the first time in the literature, the 
cross section for the production of a heavy meson molecule in ultra-peripheral 
collisions.  We have combined a effective Lagrangian to compute the amplitude 
of the process $\gamma \gamma \to D^+ D^-$ with a prescription to project this 
amplitude onto the amplitude for bound state formation. The resulting 
$\gamma \gamma \to B$ cross section was then convoluted with the equivalent 
photon fluxes from the projetile and target and the final cross section 
$\sigma_B(AA \to AAB)$ was obtained. For $\sqrt{s_{NN}} = 5.02$ TeV it  is
$3.0^{+0.8}_{-1.2} \,\, \mu b$. This number is consistent with the  results
obtained for other scalar exotic charmonium molecules in \cite{br} and 
\cite{fa}. The  parameters of the calculation are $\Lambda$, 
$\hat{q}$ and $E_b$, which are the hadronic form factor cut-off, the maximum 
momentum of an emitted photon and the binding energy, respectively. All these
parameters can be constrained by experimental information and by calculations.
Thus, we believe that in the future it will be possible to increase the 
precision of our calculation.

\begin{acknowledgments}
We are deeply indebted to K. Khemchandani and A. Martinez Torres for 
instructive discussions. 
This work was  partially financed by the Brazilian funding
agencies CNPq, CAPES, FAPESP,  FAPERGS and INCT-FNA (process number          
464898/2014-5). F.S.N.  gratefully acknowledges the  support from the 
Funda\c{c}\~ao de Amparo \`a  Pesquisa do Estado de S\~ao Paulo (FAPESP).   
C.A.B. acknowledges support by the U.S. DOE Grant DE-FG02-08ER41533 and the 
Helmholtz Research Academy Hesse for FAIR.

\end{acknowledgments}

\appendix

\section{Cross section of the process 
$ e^+ \, e^-  \to e^+ \, e^- \, D^+ \, D^-$}
\label{apa}

In this appendix we will adapt Eq. (\ref{sigmafp2}) to the 
$e^+e^-\rightarrow e^+e^-c\bar{c}$ process.  We start from Eq.(\ref{defn})
with $Z=1$:
\beq
n({\bf q}) d^3 q = \frac{\alpha}{\pi^2 \omega}
\frac{({\bf q}_{\perp})^2}{\left( ({\bf q}_{\perp})^2
+ (\omega/\gamma)^2\right)^2} d^3 q \, .
\label{defn2}
\eeq 
First we recall that $d^3 q = d q_x dq_y dq_z = 
q_{\perp} d q_{\perp} d \theta dq_z = 1/2 d q_{\perp}^2 d \theta dq_z 
\to \pi d q_{\perp}^2 dq_z$. Then we make the following change of variables: 
$$
d q_{\perp}^2 d q_z = \frac{\omega}{\sqrt{\omega^2 - q_{\perp}^2}} 
d q_{\perp}^2 d \omega  \, .
$$
After changing the variables we integrate Eq. (\ref{defn2}) over $q_{\perp}^2$:
\begin{equation}
n(\omega) = \frac{\alpha}{\pi}\int\limits_0^{\omega^2} 
\frac{q_{\perp}^2}{[q_{\perp}^2 
+ (\omega/\gamma)^2]^2}\frac{1}{\sqrt{\omega^2 - q_{\perp}^2}}
dq_{\perp}^2 \, .
\end{equation}
The solution of this integral is:
\begin{equation}
n(\omega) = 
\frac{\alpha}{\pi} \frac{1}{\omega} 
\frac{ \gamma }{ (1+ \gamma^2)^{3/2} }
\left[2\gamma^2 \text{arcsinh}(\gamma)  
+ \text{arcsinh}(\gamma)-\gamma \sqrt{1+\gamma^2}
\right] \, .
\label{newn}
\end{equation}
After these changes in Eq. (\ref{sigmafp2}), we can write the cross section for 
the process $e^+ \, e^-\rightarrow e^+ \, e^- \, D^+ \, D^-$ inserting 
Eq. (\ref{newn}) into Eq. (\ref{sigmafp2}) and recalling that  
for electrons we use $\gamma = \sqrt{s}/2 m_e$ and also $\hat{q} = m_e$.

\section{$Pb Pb$ in QED versus $pp$ in QCD} 
\label{apb}
In this paper we have been presenting predictions for quantities which are 
poorly known experimentally. In order to know, at least, what to expect and to 
have an idea of the order of magnitude of the cross sections we present below
an estimate of the cross sections of the QED process 
$Pb \, Pb \to Pb \, Pb \, c \, \bar{c} $ and of the QCD process 
$p \, p \to p \, p \, c \, \bar{c} $. The latter can be calculated with the 
simple convolution formula of the parton model:
\begin{equation}
\sigma(p \, p \rightarrow p \, p \, c \, \bar{c}) = \int 
\limits_{\frac{4m_c^2}{s}}^1 dx_1f(x_1)
\int\limits_{\frac{4m_c^2}{x_1s}}^1 dx_2f(x_2)\sigma(x_1,x_2) \, , 
\end{equation}
where $x_1$ and $x_2$ are the proton momentum fractions of the colliding 
partons.
In the above expression the integration limits come from the kinematical 
constraint $ x_1 x_2 s \geq 4m_c^2 $.
We know that this reaction is domintated by the elementary process     
$g \, g \to c \, \bar{c}$. In a rough approximation the gluon momentum 
distributions and the  elementary $\sigma(g \, g \to c \, \bar{c})$ cross 
section are given by:
\begin{equation}
f(x_1) = 1/x_1;\quad f(x_2) = 1/x_2;\quad \sigma(x_1,x_2) 
= \frac{\alpha_s^2}{x_1x_2 s} \, .
\label{defspp}
\end{equation}
With these choices the integral above can be easily performed and yields:
\begin{equation}
\sigma(pp\rightarrow ppc\bar{c}) = \frac{\alpha_s^2}{4m_c^2}
\left[\ln\left(\frac{s}{4m_c^2}\right)-\left(1-\frac{4m_c^2}{s}\right)\right] 
\, .
\label{sigpp}
\end{equation}
In the case of charm production in an UPC of Pb-Pb we have an analogous 
convolution formula written in terms of the energies $\omega_1$ and $\omega_2$ 
of the colliding photons.
Assuming that the maximum energy carried by one emitted photon is $\sqrt{s}/2$, 
the cross section is written as:
\begin{equation}
\sigma(Pb \, Pb \rightarrow Pb \, Pb \, c \, \bar{c}) 
= \int\limits_{\frac{2m_c^2}
{\sqrt{s}}}^{\frac{\sqrt{s}}{2}} 
d\omega_1n(\omega_1)\int\limits_{\frac{m_c^2}
{\omega_1}}^{\frac{\sqrt{s}}{2}} d\omega_2n(\omega_2)
\sigma(\omega_1,\omega_2) \, . 
\label{sigchuchu}
\end{equation}
In the above expression the integration limits come from the kinematical 
constraint 
$\hat{s} = (k_1+k_2)^2 = 2(k_1\cdot k_2)= 2(\omega_1\omega_2 -
(\omega_1)(-\omega_2)) = 4\omega_1\omega_2\geq 4m_c^2$.
The number of equivalent photons with energy $\omega$, $n(\omega)$, and the 
photon-photon fusion cross section into an object with invariant mass 
$4 \omega_1 \omega_2$ can be roughly approximated by
\begin{equation}
n(\omega_1) = \frac{Z^2\alpha}{\omega_1};\quad n(\omega_2) = 
\frac{Z^2\alpha}{\omega_2} ;\quad \sigma(\omega_1,\omega_2) = 
\frac{\alpha^2}{4\omega_1\omega_2} \, .
\label{defschuchu}
\end{equation}
We note the  similarity between the above expressions and 
(\ref{defspp}). As before this integral can be easily solved and we find:
\begin{equation}
\sigma(Pb \, Pb\rightarrow Pb \, Pb \, c \, \bar{c}) 
= \frac{Z^4\alpha^4}{4m_c^2}     
\left[\ln\left(\frac{s}{4m_c^2}\right)-\left(1-\frac{4m_c^2}{s}\right) 
\right] \, .
\label{sigpbpb}
\end{equation}
Not surprisingly, (\ref{sigpp}) and (\ref{sigpbpb}) are identical except for 
the pre-factors. Using $Z = 82$, $\alpha = 1/137$ and $\alpha_s^2=0.1$, we 
find
\begin{equation}
\frac{\sigma(Pb \, Pb \rightarrow Pb \, Pb \, c \, \bar{c})}
{\sigma(p \, p \rightarrow p \, p \, c \, \bar{c})} = 
\frac{Z^4\alpha^4}{\alpha_s^2} \approx \frac{0.128}{0.1} = 1.28 \, .
\label{ratio}
\end{equation}
We could assume that $\alpha_s^2=0.2$, which is also a reasonable value. Then
the above ratio would have been $0.64$. 

In Eq. (\ref{defspp}) we could improve the approximation for $f(x)$. 
At increasingly higher energies the more singular behavior of the gluon
distribution can be represented by $f(x) \simeq 1/x^{1+\delta}$, with 
$\delta \simeq 0.5$. Analogously, in Eq. (\ref{defschuchu}) we could improve 
the approximation for $n(\omega)$ including the $ln(\omega)$ correction. This 
would change both cross sections in (\ref{ratio}) in the same direction.

From this exercise we conclude that, for charm inclusive production at the 
same nucleon-nucleon center of mass energy, we have:
\begin{equation}
\sigma_{QED}(Pb \, Pb \rightarrow Pb \, Pb \, c \, \bar{c}) 
\approx \sigma_{QCD}(p \, p \rightarrow p \, p \, c \, \bar{c}) \, .
\label{theo}
\end{equation}
The above approximate identity is, of course, very crude but it tells us that
the two reactions have comparable cross sections.

\end{document}